# Observation of  Relaxation Phenomena in  Thermophysical   Properties of Alloys and Metals

Magdalena Anna Pelc

Institute of Physics Maria –Sklodowska University, Lublin, Poland

Miroslaw Kozlowski

Physics Department, Warsaw University, Warsaw, Poland

Abstract

In this paper the proposal for the study of the second sound in medium is presented. The master equation is derived and its solution is obtained, The properties of alloys  with very long relaxation times , as the examples for the proposed study are discussed

# Chapter 1

# Overview of the research

In the description of the evolution of any physical system, it is mandatory to evaluate, as accurately as possible, the order of magnitude of different characteristic time scales, since their relationship with the time scale of observation (the time during which we assume our description of the system to be valid) will determine, along with the relevant equations, the evolution pattern. Take a forced damped harmonic oscillator and consider its motion on a time scale much larger than both the damping time and the period of the forced oscillation. Then, what one observes is just a harmonic motion. Had we observed the system on a time scale of the order of (or smaller) than the damping time, the transient regime would have become apparent. This is rather general and of a very relevant interest when dealing with dissipative systems. It is our purpose here, by means of examples and arguments related to a wide class of phenomena, to emphasize the convenience of resorting to hyperbolic theories when dissipative processes, either outside the steady-state regime or when the observation time is of the order or shorter than some characteristic time of the system, are under consideration. Furthermore, as it will be mentioned below, transient phenomena may affect the way in which the system leaves the equilibrium, thereby affecting the future of the system even for time scales much larger than the relaxation time.

Parabolic theories of dissipative phenomena have long and a venerable history and proved very useful especially in the steady-state regime [1]. They exhibit however some undesirable features, such as acausality (see e.g., [2], [3]), that prompted the formulation of hyperbolic theories of dissipation to get rid of them. This was achieved at the price of extending the set of field variables by including the dissipative fluxes (heat current, non-equilibrium stresses and so on) at the same footing as the classical ones (energy densities, equilibrium pressures, etc), thereby giving rise to a set of more physically satisfactory (as they much better conform with experiments) but involved theories from the mathematical point of view. These theories have the additional advantage of being backed by statistical fluctuation theory, kinetic theory of gases (Grad's 13-moment approximation), information theory and correlated random walks (at least in the version of Jou *et al.*) [3].

A key quantity m these theories is the relaxation time $\tau$ of the corresponding dissipative process. This positive-definite quantity has a distinct physical meaning, namely the time taken by the system to return spontaneously to the steady state (whether of thermodynamic equilibrium or not) after it has been suddenly removed from it. It is, however, connected to the mean collision time $t_c$ of the particles responsible for the dissipative process It is therefore appropriate to interpret the relaxation time as the time taken by the corresponding dissipative flow to relax to its steady value. Thus, it is well known that the classical Fourier law for heat current,

$$\vec{q} = -\kappa \vec{\nabla} T$$

(1.1) with $\kappa$ the heat conductivity of the fluid, leads to a parabolic equation for temperature (diffusion equation)

$$\frac{\partial T}{\partial t} = \chi \nabla^2 T, \qquad \chi = \frac{\kappa}{\rho C_p}$$

(1.2)

which does not forecast propagation of perturbations along characteristic causal light-cones That is to say, perturbations propagate with infinite speed. This non-causal behavior is easily visualized by taking a look at the thermal conduction in an infinite one dimensional medium (see e.g. [1], [8]). Assuming that the temperature of the line is zero for $t < 0$, and putting a heat source at $x = x_Q$ when t = O, the temperature profile for $t >$ is given by

$$T \sim \frac{1}{\sqrt{t}} \exp\left[-\frac{(x-x_0)^2}{t}\right]$$

The origin of this behavior can be traced to the parabolic character of Fourier's law, which implies that the heat flow starts (vanishes) simultaneously with the appearance (disappearance) of a temperature gradient. Although $\tau$ is very small for phonon-electron, and phonon-phonon interaction at room temperature neglecting it is the source of difficulties, and in some cases a bad approximation as for example in superfluid Helium [10], and degenerate stars where thermal conduction is dominated by electrons -see [3], [5], [11], for further examples.

In order to overcome this problem Cattaneo and (independently) Vernotte by using the relaxation time approximation to Boltzmann equation for a simple gas derived a generalization of Fourier's law, namely [12]

$$\tau \frac{\partial \vec{q}}{\partial t} + \vec{q} = -\kappa \nabla T$$

This expression (known as Cattaneo-Vernotte's equation) leads to a hyperbolic equation for the temperature (Heaviside equation) which describes the propagation of thermal signals with a finite speed

$$v = \sqrt{\chi/\tau}$$

(1.5) This diverges only if the unphysical assumption of setting $\tau$ to zero is made.

It is worth mentioning that a simple random walk analysis of transport processes naturally leads to Heaviside equation, not to the diffusion equation Again, the latter is obtained only if one neglects the second derivative term.

It is instructive to write (1.4) in the equivalent integral form

$$\vec{q} = -\frac{\chi}{\tau} \int_{-\infty}^{t} \exp\left[ -\frac{(t-t')}{\tau} \right] \cdot \nabla T(x',t') dt'$$

(1.6) which is a particular case of the more general expression

$$\vec{q} = -\int_{-\infty}^{t} Q(t-t') \nabla T(x',t') dt'$$

(1.7)

The physical meaning of the kernel $Q\,(t - t'')$ becomes obvious by observing that

$$\text{for} \quad Q = \kappa \delta(t-t') \rightarrow \vec{q} = -\kappa \nabla T \tag{1.8}$$

$$\text{for} \quad Q = \text{constant} \rightarrow \frac{\partial^2 T}{\partial t^2} = \chi \nabla^2 T$$

i.e., $Q$ describes the thermal memory of the material by assigning different weights to temperature gradients at different moments in the past. The Fourier law corresponds to a zero-memory material (the only relevant temperature gradient is the "last" one, i.e., the one simultaneous with the appearance of $q$). By contrast the infinite memory case (with $Q$ = constant) leads to an undamped wave. Somewhere in the middle is the Cattaneo-Vernotte equation, for which all temperature gradients contribute to $q$, but their relevance goes down as we move to the past.

From these comments it should be clear that different classes of dissipative systems may be described by different kernels. The one corresponding to (1.4) being suitable for the description of a restricted subclass of phenomena.

Obviously, when studying transient regimes, i.e., the evolution from a steady-state situation to a new one, $\tau$ cannot be neglected. In fact, leaving aside that parabolic theories are necessarily non-causal, it is obvious that whenever the time scale of the problem under consideration becomes of the order of (or smaller) than the relaxation time, the latter cannot be ignored. It is common sense what is at stake here: neglecting the relaxation time amounts -in this situation- to disregarding the whole problem under consideration. According to a basic assumption underlying the disposal of hyperbolic dissipative theories, dissipative processes with relaxation times comparable to the characteristic time of the system are out of the hydrodynamic regime. However, the concept of hydrodynamic regime involves the ratio between the mean free path of fluid particles and the characteristic length of the system. When this ratio is lower that unity, the fluid is within the hydrodynamic regime. When it is larger than unity, the regime becomes Knudsen's. In the latter case the fluid is no longer a continuum and even hyperbolic theories cease to be realiable.

Therefore that assumption can be valid only if the particles making up the fluid are the same ones that transport the heat. However, this is (almost?) never the case. Specifically, for a neutron star, $r$ is of the order of the scattering time between electrons (which carry the heat) but this fact is not an obstacle (no matter how large the mean free path of these electrons may be) to consider the neutron star as formed by a Fermi fluid of degenerate neutrons. The same is true for the second sound in superfluid Helium and solids, and for almost any ordinary fluid. In brief, the hydrodynamic regime refers to fluid particles that not necessarily (and as a matter of fact, almost never) transport the heat. Therefore large relaxation times (large mean free paths of particles involved in heat transport) does not imply a departure from the hydrodynamic regime, but it is usually overlooked).

However, even in the case that the particles making up the fluid are responsible of the dissipative process, it is not "always" valid to take for granted that $\tau$ and $t_c$ are of the same order  or what comes to the same that the dimensionless quantity $T = (\tau c_s/L)^2$ is negligible in all instances -here $c_s$ stands for the adiabatic speed of sound in the fluid under consideration and $L$ the characteristic length of the system. That assumption would be right if $T$ were always comparable to $t_c$ and $L$ always "large", but there are, however, important situations in which $\tau >\bullet t_c >$ and $L$ "small" although still large enough to justify a macroscopic description. For tiny semiconductor pieces of about $10\sim^4$ cm in size, used in common electronic devices submitted to high electric fields, the above dimensionless combination (with $\tau \sim 10\sim^{10}$ sec, $c_s \sim 10^7$ cm/sec [17]) can easily be of the order of unity. In ultrasound propagation as well as light-scattering experiments in gases and neutron-scattering in liquids the relevant length is no longer the system size, but the wave  lenght A which is usually much smaller than $L$  Because of this, hyperbolic theories may bear some importance in the study of nanoparticles and quantum dots. Likewise m polymeric fluids relaxation times are related to the internal configurational degrees of freedom and so much longer than $t_c$ (in fact they are in the range of the minutes), and $c_a \sim 10^5$ cm/sec, In the degenerate core of aged stars the thermal relaxation time can be as high as l second [20]. Assuming the radius of the core of about $10\sim^2$ times the solar radius, one has F $\sim O(l)$ again. Fully ionized plasmas exhibit a collisionless regime (Vlasov regime) for which the parabolic hydrodynamics predicts a plasmon dispersion relation at variance with the microscopic results; the latter agree, however, with the hyperbolic hydrodynamic approach. Think for instance of some syrup fluid flowing under a imposed shear stress, and imagine that the shear is suddenly switched off. This liquid will come to a halt only after a much longer time ($\tau$) than the collision time between its constituent particles has elapsed.

Even in the steady regime the descriptions offered by parabolic and hyperbolic theories do not necessarily coincide. The differences between them in such a situation arise from (i) the presence of $\tau$ in terms that couple the vorticity to the heat flux and shear stresses. These may be large even in steady states (e.g. rotating stars). There are also other acceleration coupling terms to bulk and shear stresses and heat flux. The coefficients for these vanish in parabolic theories, and they could be large even in the steady state. (ii) From the convective part of the time derivative (which are not negligible in the presence of large spatial gradients). (iii) from modifications in the equations of state due to the presence of dissipative fluxes.

Chapter 2

# Thermophysical properties of the alloys and metals

## 1. INTRODUCTION

Relaxation phenomena in thermophysical properties occur after changes in temperature. Opportunities for observations of such phenomena are provided by subsecond thermophysical measurements [1-4]. Under usual conditions, temperature changes slowly, which permits establishment of equilibrium. This means that temperature changes are too slow to observe fast relaxation processes. An example of a very fast process is electron-phonon relaxation in metals. When the electrons are heated above the temperature of the lattice, the relaxation time is of the order of $10^{-12}$s. However, it becomes measurable when the sample is heated rapidly. For instance, electron-phonon relaxation in a thin niobium film (about 20 nm) has been studied [5]. The electrons were excited with a laser pulse having a duration of the order of $10^{-13}$ s. Changes in the transmitted probe intensity were measured for time delays up to $1.5 \times 10^{-12}$ s with respect to the pump. The probe signal depended on the lattice temperature, so that relaxation was clearly seen. In another investigation [6], the transient reflectivity and transmittivity of thin gold films were measured. The observed relaxation was interpreted as a result of thermalization of the electron gas. The thermalization time, of the order of $10^{-12}$ s, is thus comparable to the electron-phonon time.

On the other hand, processes are known for which relaxation times are sufficiently long to make the phenomena easily observable. Relaxation phenomena appear in diffusion-controlled processes, such as ordering in alloys and formation of point defects in the lattice of crystals. Studies of relaxation provide information not available from measurements under equilibrium conditions. First, the kinetics of the equilibration can be determined. Second, and sometimes even more important, the observations permit a reliable separation of contributions to physical properties of metals. They become distinguishable due to different relaxation times. The equilibration can be monitored through various physical properties of the sample. The simplest approach is to measure its electrical resistivity, but usually changes in the resistivity are relatively small. Observations of the relaxation in enthalpy and specific heat are important because they yield energetic characteristics of the process under study.

In this brief review, the following phenomena related to relaxation are considered: (i) ordering in alloys, (ii) glass transition in supercooled liquids, and (iii) equilibration of point defects. The main emphasis is placed on observations of the equilibration of point defects. At present, they play a major role in the determination of equilibrium concentrations of point defects and their influence on properties of metals at high temperatures. An important method of studying vacancy formation is based on rapid cooling (quenching) of the samples from high temperatures. The quenched-in vacancies become immobile at low temperatures, so that this nonequi-librium state can be held for a long time. Usually, it is monitored by measurements of the extra resistivity caused by the vacancies. The aim of Quenching experiments is twofold: (i) to reveal the vacancy concentrations at high temperatures and the formation enthalpies and (ii) to determine the migration enthalpies through annealing of the quenched samples.

The methods of studying relaxation in thermophysical properties of metals may be grouped as follows.

(a)    The sample is rapidly heated to a higher temperature (or cooled to a lower temperature), and the equilibration is monitored through measurements of a proper physical property of the sample, e.g., electrical resistivity, enthalpy, or the parameters of positron annihilation. This method has the advantage that both initial and final states of the sample are well defined. As a drawback, relatively small changes in the chosen property have to be measured. This technique was employed in studies of ordering in alloys [7-10] and in measurements of enthalpy related to equilibrium vacancies [11]. Also, the vacancy equilibration was studied with the positron annihilation [12]. Owing to the high sensitivity of the method, the measurements were carried out far below the melting point, so that the equilibration time was sufficiently long. With the resistivity as a probe, the measurements should be performed at higher temperatures and hence shorter relaxation times.

(b)   The sample is rapidly heated to a high temperature, kept at this temperature for an adjustable time, and then quenched. Resistivity after quenching is measured and related to the time of exposure to high tem-perature. This method was employed in studies of the vacancy equilibration [13-15]. However, it was difficult to evaluate correctly the vacancy contribution at high temperatures. During quenching, there is sufficient time for many vacancies to be annihilated or to form clusters. When the quenched vacancies form clusters, their contribution to resistivity decreases. The extra resistivity of quenched samples becomes therefore smaller than the vacancy contribution at the high temperature. An important advantage of this approach is the high sensitivity. The extra resistivity is measured at liquid helium temperatures, where it makes the main contribution. For instance, an extra resistivity of 1 ppm of the total resistivity at high temperatures is measurable at low temperatures. The vacancy equilibration is thus observable even at medium temperatures where the extra resistivity is small but the relaxation time is sufficiently long.

The sample is subjected to such rapid oscillations in temperature that equilibration cannot follow them. This technique was used in measurements of the specific heat of supercooled liquids on approaching the glass transition [16-18] and in observations of the vacancy equilibration [19-23]. Under rapid temperature oscillations, the influence of the vacancies is almost completely excluded. This relates to properties that depend on changes of the vacancy concentrations during the measurements: specific heat, thermal expansivity, and temperature derivative of resistivity. If the vacancy concentration does not follow the temperature oscillations and retains a mean value, then these properties practically correspond to an ideal defect-free crystal. This method permits a reliable separation of vacancy contributions to the physical properties. A drawback of this approach arises from short equilibration times due to the high mobility of the vacancies at high temperatures and numerous internal sources (sinks) for them. Relaxation is therefore observable only at high frequencies of the temperature oscillations. The amplitude of the temperature oscillations is inversely proportional to their frequency, and the measurements require employment of special techniques.

## 2. ORDERING IN ALLOYS

Measurements of electrical resistivity are well-known as an effective tool for investigations of ordering in alloys. For example, the kinetics of the order-disorder transition in $Cu_3$ Au alloy was studied through the electrical resistivity [7]. The relaxation time from the disordered state to the equilibrium state below the transition point is so long that the measurements can be carried out very easily. The sample was heated in a furnace at 673 K and then quenched to and kept at a certain temperature below the order-disorder transition, 664.2 K. The relaxation time increases rapidly when the temperature approaches the transition point

The kinetics of long-range ordering in a $Ni_3Al$ compound was studied through measurements of resistivity [8]. The sample annealed at 1173 K was rapidly cooled to lower temperatures. An exponential decrease in resistivity at 862 and 880 K and increase at 925 and 973 K were observed. The inverted resistivity relaxation is in agreement with a phenomenological model [24,25]. The observed changes in the resistivity never exceeded 0.2% and the uncertainty in the relaxation times was estimated as 30%. The relaxation time in the range 862 to 973 K decreases from about $10^6$ to $2 \times 10^5$ s.

A similar study of the short-range ordering in Au-Ag alloys has been reported [9]. When a constant value of resistivity was achieved, a small change in temperature, typically 10 K, led to another equilibrium value for the resistivity. The changes of the resistivity were of the order of 0.1 %. In the range 450 to 520 K, the relaxation time decreases from about $10^5$ to $10^2$s, also being dependent on the Ag contents. The activation enthalpy has been determined from the Arrhenius plot of the relaxation time. In contrast to the previous results, the relaxation did not obey exactly a pure exponential law.

An investigation of resistivity, along with measurements of the velocity of sound, has been performed in amorphous alloys FeCrB, FeMnB, and FeNiB [10]. In metallic glasses, structural relaxation occurs at temperatures sufficiently high to allow appreciable atomic motion but low enough to avoid crystallization. A slow relaxation of the resistivity of both signs was observed, being dependent on the composition and temperature.

In all the above examples, the relaxation times were very long. However, they rapidly decrease with increasing temperature. Subsecond thermophysical measurements may therefore become useful for studying ordering and structural relaxation at high temperatures.

**3. GLASS TRANSITION IN SUPERCOOLED LIQUIDS**

When a liquid is supercooled sufficiently below its equilibrium freezing point, it inevitably undergoes a glass transition into a state with thermodynamic properties appropriate for a solid. The characteristic relaxation times of the liquid increase rapidly as the transition point is approached. The specific heat of a supercooled liquid is greater than that of the crystal. The temperature at which the specific heat changes abruptly depends on the cooling or heating rate, while the specific heat of the liquid measured with the modulation calorimetry becomes frequency dependent.

The method, known as the third-harmonic technique, has been developed by Corbino [26], the founder of modulation calorimetry. A bridge circuit is employed to reduce the large signal of the fundamental frequency, and the third harmonic is measured with a lock-in amplifier. Temperature oscillations in the heater depend on the product of C, the specific heat, and $k$, the thermal conductivity of the liquid surrounding the heater. The measurements covered a frequency range of five decades. When the period of the temperature oscillations becomes comparable with the relaxation time, the quantity $Ck$ contains both real and imaginary parts. The frequency corresponding to the change of the real part and the peak of the imaginary part depends strongly on the temperature. It has been shown that the frequency dependence pertains only to the specific heat. The relaxation time obtained in specific-heat measurements has the same temperature dependence as that from other techniques.

Relaxation phenomena can also be studied through observations of the phase shift between the temperature oscillations and the corresponding changes in the chosen property of the sample. A modulation calorimeter has been reported [27] for a wide frequency range, also employing the third-harmonic technique.

## 4. VACANCY EQUILIBRATION IN METALS

Observations of the vacancy equilibration are considered the most reliable method to separate vacancy contributions to physical properties of metals. The relaxation time, i.e., the characteristic time of setting up the equilibrium concentration, depends not only on the temperature and migration parameters but also on the density of internal sources (sinks) for the vacancies. The relaxation time, $\tau$, is proportional to the square of the mean distance $L$ between the sources (sinks) and inversely proportional to the coefficient of self-diffusion of the vacancies, $D_v$: $\tau = AL^2/D_v$ where $A$ is a numerical coefficient depending on the geometry of the sources (sinks). Studies of the vacancy equilibration provide data for calculations of both formation and migration enthalpies of the vacancies.

The vacancy equilibration in gold, platinum, and aluminum was observed through the extra resistivity of the samples subjected to controllable exposures to high temperatures. In gold [13], two methods were used. First, thin single-crystal slabs were heated to 1150-1200 K with a rapid jet of hot compressed air, held for a short period of time, and then quenched. The frozen-in vacancies were detected by precipitating them as vacancy tetrahedra observable by a transmission electron microscope. Second, polycrystalline foils were pulse heated resistively to a high temperature, held at this temperature for durations ranging from 0.02 to 1.5 s, and then quenched. The extra resistivity after quenching was measured at 4.2 K. The half-times for the vacancy equilibration at 926 and 1151 K have been determined as 80 and 9.5 ms, respectively. The results indicated that free dislocations were the predominant vacancy sources in the samples. In studies on platinum [14], a wire sample was heated to a quench temperature by the discharge current of a capacitor at a rate of about $10^6 Ks^{-1}$. Cool helium gas blown across the wire provided a rapid quench after termination of the heating current. For measuring the extra resistivity, the sample was immersed in liquid nitrogen or helium. The same technique was employed in studies on aluminum [15]. The extra resistivity due to vacancies in terms of the pulse-heating time was measured at temperatures from 583 to 673 K. The vacancy contribution to the resistivity and the migration enthalpy were calculated.

In another study [28], the vacancy equilibration in tungsten was monitored by interrupting quenches from 2900 K at well-defined temperatures from 1550 to 2600 K, followed by a quench to 300 K. The quenched-in resistivity was measured at 4 K. From the data obtained, migration enthalpy was found to increase from 1.68 eV at 1550 K to 2.02 eV at 2600 K. A similar temperature dependence of the formation enthalpy could explain the observed curvature in the Arrhenius plot of self-diffusion in tungsten.

The heat absorbed by the vacancy formation in aluminum was measured directly with a precision microcalorimeter [11]. The temperature of the calorimetry chamber was stabilized and the sample, held at a different temperature, was lowered into the thermopile. In the absence of thermal reactions in the sample, the output voltage of the thermopile varies exponentially. This was checked with a copper sample in which the vacancy contributions at these temperatures are negligible. If extra heat is absorbed or released by the sample, the output voltage no longer behaves exponentially. The extra heat was attributed to changes in the vacancy concentrations in the sample. The investigators believed that their results were underestimated because the initial part of the calorimetric curves could not be followed.

The vacancy contribution to the specific heat of refractory metals can be seen from the results of rapid-heating measurements. Specific heat values under rapid-heating conditions lie below equilibrium data at intermediate temperatures but increase rapidly when temperature approaches the melting point [2]. Such a behavior should be expected when the heating rate is high enough to avoid vacancy formation at intermediate temperatures but becomes insufficient at higher temperatures. If this interpretation is correct, then the total vacancy contribution to the enthalpy in such experiments should be the same as in equilibrium measurements.

## 5. PROPOSAL FOR SUBSECOND THERMOPHYSICS

Rapid-heating experiments may reveal the vacancy contributions to the entalpy of metals. Under very rapid heating, the vacancies have no time to appear, so that the enthalpy at a given premelting temperature should be smaller than in equilibrium measurements. For molybdenum, the expected difference based on the supposed vacancy origin of the nonlinear increase in the specific heat amounts to about 10%. To check this assumption, an examination of typical data on molybdenum has been undertaken. They include equilibrium measurements of the enthalpy [29], millisecond-resolution measurements of the specific heat [30-33], and rapid-heating determinations of the enthalpy at the melting point [34-37]. Taking into account an estimated time of vacancy equilibration, only experiments with heating rates of the order of $10^8$ K s$^{-1}$ or more may be assumed not to include the vacancy contribution. Rapid-heating experiments usually start at room temperatures. To fit the specific-heat data obtained at high temperatures, the enthalpy at 1500 K was taken as 33.5kJmol$^{-1}$ [29]. Certainly different values of the enthalpy at the melting point under equilibrium and rapid-heating conditions appear in the literature. To make a quantitative comparison, parts of the enthalpy related to the nonlinear increase in specific heat were separated from the results of equilibrium measurements. For this purpose, the experimental data were fitted by equations taking into account the vacancy contributions.

Thus we have three sets of data: (i) equilibrium enthalpies at the melting points after subtracting the supposed vacancy contributions, $H_1$;

(ii) total equilibrium enthalpies, $H_2$; (iii) enthalpies from rapid-heating measurements, $H_3$, which are expected to be close to $H_x$ rather than to $H_2$. The difference between $H_x$ and $H_2$ is therefore quite detectable. At present, the rapid-heating data, $H_3$, lie between the two values. The reasons may be as follows: (i) the heating is not fast enough to avoid vacancy formation completely, (ii) the superheat of the samples under rapid heating leads to higher enthalpies at the apparent melting point, and (iii) vacancy formation accounts for only a part of the nonlinear increase in specific heat. To check these possibilities, it is enough to vary the heating rates and determine the apparent melting point or to measure the enthalpy at a selected premelting temperature.

To determine vacancy contributions to enthalpy of metals at high temperatures, a more convenient and straightforward approach can be proposed. After heating the sample to a premelting temperature, the initial cooling curve depends on whether the vacancies had time to arise. If they did not, they will appear immediately after heating. At premelting temperatures, equilibrium vacancy concentrations set up in $10^{-4}$ to $10^{-2}$ s in low-melting point metals and in $10^{-8}$ to $10^{-6}$ s in refractory metals [23]. Under usual conditions, the temperature of the sample after heating, in the time region of interest, remains nearly constant. Therefore, heat absorbed by the vacancy formation after heating should be easily detected. The shape of the cooling curve depends on the vacancy contribution to the enthalpy and on the relaxation time, which depends strongly on the temperature. For molybdenum, the expected decrease is of about 200 K. If heating is not sufficiently fast, the phenomenon could be studied under gradually increasing the upper temperature of the sample. The decrease in the temperature immediately after heating must first increase with the temperature, reach a maximum, and then fall due to the decrease in relaxation time. Also, it is easy to show that the decrease in the temperature of the sample after rapid heating should be accompanied by an increase in its volume. Such unusual behavior would be the best confirmation of the vacancy origin of the phenomenon. The proposed approach seems to be the simplest experiment that could be performed for the determination of equilibrium vacancy concentrations in metals.

**REFERENCES**

1. A. Cezairliyan, in *Compendium of Thermophysical Property Measurement Methods,* Vol. 1, K. D. Maglic, A. Cezairliyan, and V. E. Peletsky, eds. (Plenum Press, New York, 1984), pp. 643-668.

2. S. V. Lebedev and A. I. Sawatimskii, *Sov. Phys. Usp. 21:149* (1984).

3. G. R. Gathers, *Rep. Prog Phys.* 49:341 (1986).

4. A. Cezairliyan, G. R. Gathers, A. M. Malvezzi, A. P. Miiller, F. Righini, and J. W. Shaner, *Int. J. Thermophys.* **11:819** (1990).

5. M. Mihailidi, Q. Xing, K. M. Yoo, and R. R. Alfano, *Phys. Rev. B* 49:3207 (1994).

6. C.-K. Sun, F. Vallee, L. H. Acioli, E. P. Ippen, and J. G. Fujimoto, *Phys. Rev. B* 50:15337 (1994).

7. T. Hashimoto, T. Miyoshi, and H. Ohtsuka, *Phys. Rev. B* 13:1119 (1976).

8. R. Kozubski and M. C. Cadeville, /. *Phys. F Metal Phys.* 18:2569 (1988).

9. T. Doppler and W. Pfeiler, *Phys. Stat. Solidi A* 131:131 (1992).

10. G. Riontino, G. W. Koebrugge, M. Baricco, and J. Sietsma, *Phys. Stat. Solidi B* 179:315 (1993).

11. G. Guarini and G. M. Schiavini, *Phil. Mag.* **14:41** (1966).

12. H.-E. Schaefer and G. Schmid, /. *Phys. Condens. Matter* 1 (Suppl. A):SA49 (1989).

13. D. N. Seidman and R. W. Balluffi, *Phys. Rev.* 139:A1824 (1965).

14. F. Heigl and R. Sizmann, *Crystal Lattice Defects* 3:13 (1972).

15. K. Ono and T. Kino, /. *Phys. Soc. Japan* 44:875 (1978).

16. N. O. Birge and S. R. Nagel, *Phys. Rev. Lett.* 54:2674 (1985).

17. N. O. Birge, *Phys. Rev. B* 34:1631 (1986).

18. N. O. Birge and S. R. Nagel, *Rev. Sci. Instrum.* 58:1464 (1987).

19. D. A. Skelskey and J. Van den Sype, *Solid State Commun.* 15:1257 (1974).

20. A. H. Seville, *Phys. Stat. Solidi A* 21:649 (1974).

# Chapter 3

# Second sound in solids, microscopic desription

1. Theory

Let us consider the propagation of the heat in a medium. We will describe the medium as a Fermi gas of electrons. When the medium is subjected to a heat pulse, e.g. during the laser interaction the propagation of the pulse inside the medium can be described by a nonlocal Fourier law [1]

$$\vec{q} = -\int_{-\infty}^{t} K(t-t') \nabla T(x',t') dt'$$

(1)

where $q(t)$ denotes energy flux density, $T$ is the temperature and $K(t\text{-}t')$ denotes the memory function for thermal processes. The energy flux density $q(t)$ fulfils the nonlocal heat transport equation:

$$\frac{\partial}{\partial t} T(t) = \frac{1}{\rho\, c_V} \nabla^2 \int_{-\infty}^{t} K(t-t') T(t') dt',$$

(2)

where $\rho$ is the density of Fermi gas of electrons and $C_V$ is the specific heat at constant volume.

For a system with very short memory

$$K(t-t') = K_1 \delta(t-t').$$

(3)

we obtain from eq.(2)

$$\frac{\partial}{\partial t} T(t) = \frac{1}{\rho\, c_V} K_1 \nabla^2 T.$$

(4)

By comparing eq.(4) with the kinetic theory formulation of heat transport [8] we obtain for $K_1$

$$K_1 = D c_V \rho,$$

(5)

where $D$ is the diffusivity of the nuclear medium and $C_V$ is the specific heat at constant volume, $\rho$ denotes the density of nuclear matter.

Next, we consider a system with a very long memory function, i.e.

$$K(t - t') = K_2.$$

(6)

After substituting formula (6) into formula (2) we obtain

$$\frac{\partial}{\partial t} T = \frac{K_2}{\rho \, c_V} \nabla^2 \int_{-\infty}^{t} T(t') dt,$$

(7)

and finally

$$\frac{\partial^2 T}{\partial t^2} = \frac{K_2}{\rho \, c_V} \nabla^2 T.$$

(8)

In the following, we consider the intermediate form of the memory function:

$$K(t - t') = \frac{K_3}{\tau} \exp\left[ -\frac{(t - t')}{\tau} \right],$$

(9)

where $\tau$ is the relaxation time for thermal phenomena. After substituting formula (9) into eq. (2) we obtain

$$c_V \frac{\partial^2 T}{\partial t^2} + \frac{c_V}{\tau} \frac{\partial T}{\partial t} = \frac{K_3}{\rho \tau} \nabla^2 T,$$

(10)

where

$$K_3 = D c_V \rho$$

(11)

as in formula (5). Considering formula (11) we obtain from formula (10)

$$\frac{\partial^2 T}{\partial t^2} + \frac{1}{\tau}\frac{\partial T}{\partial t} = \frac{D}{\tau}\nabla^2 T.$$

(12)

For a gas of nucleons we have for the diffusivity [2]

$$D = \frac{1}{3}v_F^3\tau,$$

(13)

where $v_F$ is the Fermi velocity. Substituting formula (13) into formula (12) we obtain the following general equation for the second sound in Fermi gas

$$\frac{\partial^2 T}{\partial t^2} + \frac{1}{\tau}\frac{\partial T}{\partial t} = \frac{1}{3}v_F^3\nabla^2 T.$$

(14)

Let us introduce the propagation velocity, $s$, for the heat pulse

$$s = \sqrt{\frac{1}{3}}v_F,$$

(15)

then eq. (14) has the form:

$$\frac{1}{s^2}\frac{\partial^2 T}{\partial t^2} + \frac{1}{\tau s^2}\frac{\partial T}{\partial t} = \nabla^2 T.$$

(16)

It is interesting to observe that for Fermi liquids one can define the sound velocity as [3]

$$v_S = \left(\frac{p_F^2}{3mm*}\left(1 + F_0^{\,S}\right)\right)^{1/2},$$

(17)

where $m$ is the mass of the free fermion (e.g. the electron), $m^*$ denotes the effective mass of interacting fermions and $F_0^{\,S}$ is the dimensionless measure of the interaction strength in the Fermi system. In the limit of weak interactions, $m* \to m$, $F_0^{\,S} \to 0$ The velocity of sound then tends towards $\sqrt{\frac{1}{3}}v_F$. In that case we observe for $v_s$,

$$v_S \to s$$

(18)

and eq. (16) can be written in the form

$$\frac{1}{\upsilon_S^2}\frac{\partial^2 T}{\partial t^2}+\frac{1}{\tau\upsilon_S^2}\frac{\partial T}{\partial t}=\nabla^2 T.$$

(19)

Eq. (19) is the damped wave equation for the propagation of the second sound pulse with propagation velocity $\upsilon_s$, which is the velocity of second sound in a Fermi gas. Equation (19) can be written as

$$\frac{1}{\upsilon_S^2}\frac{\partial^2 T}{\partial t^2}+\frac{1}{D}\frac{\partial T}{\partial t}=\nabla^2 T$$

(20)

## 2. The second sound propagation in metals

Consider a cylindrical sample of medium with unit area which is heated at one end. The temperature at the other end of the sample is detected as a function of time. The purpose of this section is to discuss the solution of eq.(19) which will be relate the signal at the temperature pulse detector (TP) $T(l, t)$ to the input pulse $T(0, t)$. The solution of eq.(19) for a cylinder of infinite length is given by

$$T(x,t)=\frac{1}{2\upsilon_S}\int dx' T(x',0)\left[\begin{array}{l} e^{-\frac{1}{2}\tau}\frac{1}{t_0}\delta(t-t_0) \\ +e^{-\frac{1}{2}\tau}\frac{1}{2\tau}\left\{\begin{array}{l} I_0\left(\frac{\left(t^2-t_0^2\right)^{\frac{1}{2}}}{2\tau}\right) \\ +\frac{t}{\left(t^2-t_0^2\right)^{\frac{1}{2}}}I_1\left(\frac{\left(t^2-t_0^2\right)^{\frac{1}{2}}}{2\tau}\right) \end{array}\right\}\Theta(t-t_0) \end{array}\right]$$

(21)

where $t_0=(x-x')/\upsilon_S$ and $I_0$ and $I_1$ are modified Bessel functions. We are concerned with the solution to eq.(20) when a nearly delta-function temperature pulse heats one end of the sample. Then at $t \sim 0$ the temperature distribution in the sample is

$$T(x,t) = \begin{cases} \Delta T_0 & \text{for } 0 < x < v_S \Delta t, \\ 0 & \text{for } x > v_S \Delta t. \end{cases}$$

With this $t = 0$ temperature profile, eq.(19) yields

$$T(l,t) = \tfrac{1}{2}\Delta T_0 e^{-t/2\tau}\Theta(t-t_0)\Theta(t_0+\Delta t-t)$$
$$+ \tfrac{1}{4}\Delta t \Delta T_0 e^{-t/2\tau}\left\{ I_0(z) + \frac{t}{2\tau}\frac{1}{z}I_1(z) \right\}\Theta(t-t_0),$$

(25)

where $z = \left(t^2 - t_0^2\right)^{1/2}/2\tau$ and $t_0 = /v_S$. The first term in this solution corresponds to ballistic propagation of the second sound damped by $\exp(-t/2\tau)$ across the sample. The second term corresponds to the propagation of the energy scattered out of the ballistic pulse by diffusion. In the limit $\tau \to \infty$ the ballistic pulse alone arrives at the detector. In the limit $\tau \to 0$ the ballistic pulse is completely damped and the second term takes an asymptotic form which is the solution to the conventional diffusion equation. In particular in this limit we have $z \to \infty$

$$I_0(z) \sim \frac{e^z}{(2\pi z)^{1/2}}, \qquad I_1(z) \sim \frac{e^z}{(2\pi z)^{1/2}},$$

(26)

so that the second term in eq.(25) becomes

$$T(l,t) \sim 2\frac{\Delta t}{4\tau}\Delta T_0 \frac{e^{-t/2\tau}\exp\left[-(t^2-t_0^2)^{1/2}/2\tau\right]}{\left[(\pi/\tau)(t^2-t_0^2)^{1/2}\right]^{1/2}}.$$

(27)

Now, for $t >> t_0$, we can write $\left(t^2 - t_0^2\right)^{1/2} \sim t - \tfrac{1}{2}\left(t_0^2/t\right)$ and thus obtain

$$\lim_{\substack{t >> t_0 \\ \tau \to 0}} T(l,t) = \Delta T_0 \frac{\Delta t}{(4\pi\tau t)^{1/2}}\exp\left(-\frac{t_0^2}{4t\tau}\right).$$

(28)

The solution to eq.(21) when there are reflecting boundaries on the sample is the superposition of the temperature at $l$ from the heated end and from image heat sources at $\pm 2nl$. This solution is

$$T(l,t) = \sum_{i=1}^{\infty} \left[ \begin{array}{l} \Delta T_0 e^{-t/2\tau} \Theta(t-t_i)\Theta(t_i+\Delta t-t) \\ + \Delta T_0 \dfrac{\Delta t}{2\tau} e^{-t/2\tau} \left\{ I_0(z_i) + \dfrac{t}{2\tau}\dfrac{1}{z_i}I_1(z_i) \right\} \Theta(t-t_i) \end{array} \right], \quad (29)$$

$t_i = t_0, 3t_0, 5t_0, \ldots = l/v_S$. For short times only the $i = 0$ term in the sum is important. Each source begins to contribute at $x = l$ at the time (1) proportional to its distance (in the ballistic limit) or (2) proportional to the square of its distance (in the diffusion limit). After sufficiently long times one obtains always the diffusion limit. At the time $t_1 \gg t_0$ all sources at a distance greater than $l_1$ given by

$$\left( \frac{l_1}{v_S \tau} \right)^2 \tau = \frac{(t_1)^2}{\tau}$$

(30)

do not contribute to the temperature at $l$ at $t_1$. Each source which contributes to the temperature at $l$ contributes the same amplitude

$$A_i = \Delta T_0 \frac{\Delta t}{(\pi \tau t_1)^{1/2}},$$

(31)

so that the total heat at $l$ is

$$T(l,t) = \sum_{i=1}^{N} \frac{\Delta T_0 \Delta t}{(\pi \tau t_1)^{1/2}},$$

(32)

where the number of contributing sources $N$ is given by $(2N+1)l = l_1$ or $N \sim (v_S/l)\left[\frac{1}{2}(\tau t_1)\right]^{1/2}$ so that eq.(32) becomes

$$T(l,t) \sim N \frac{\Delta T_0 \Delta t}{(\pi \tau t_1)^{1/2}} \sim \left( \frac{v_S \Delta t}{l} \right) \Delta T_0,$$

(33)

which is proportional to the ration of the volume of the sample initially heated by the temperature source to the total volume.

In the diffusion limit $\tau \to 0$, $t \gg t_0$ the asymptotic form of eq.(29) is

$$T(l,t) = \sum_{i=1}^{N} \frac{\Delta T_0 \Delta t}{(4\pi \tau t)^{1/2}} \exp\left(-t_i^2/4\tau t\right),$$

(34)

where the expansion of $z_i$ in eq.(29) is only valid for $t \gg t_i$. Since the $i$th term contributes to $T(l,t)$ only for $t \sim t_i^2 / \tau$, it is valid to use the asymptotic expansion of each term. Eq.(34) is the solution of the diffusion equation (19) with $\tau = 0$, i.e.

$$\frac{\partial T}{\partial t} = \tau v_S^2 \nabla^2 T.$$

(35)

This equation may also be solved in the form

$$T(x,t) = \frac{1}{l}\int_0^l T(x,0)dx + \frac{2}{l}\sum_{n=1}^{\infty}\exp\left(-n^2\pi^2\tau t / t_0^2\right) \times \cos\frac{n\pi x}{l}\int_0^l T(x',0)\cos\frac{n\pi x'}{l}dx',$$

(36)

where we have used the set of functions which naturally incorporate the boundary condition at the sample ends. For an initial temperature distribution such as (25) one finds [6]:

$$T(l,t) = \Delta T_0 \frac{v_S \Delta t}{l}\left[1 + 2\sum_{n=1}^{\infty}(-1)^n \exp\left(-n^2\pi^2\tau t / t_0^2\right)\right].$$

(37)

Note the prefactor is just the final temperature given by eq.(33). In this form it is particularly convenient to see how to use measurements of $T(l,t)$ to learn the relaxation time $\tau$. At long times, eq.(38) is asymptotically:

$$T(l,t) \sim T_f\left[1 - 2\exp\left(-\pi\tau t / t_0^2\right)\right]$$

(38)

So that $T_f(l) - T(l,t)$ may be used to measure $\tau$.

Following method presented in [1] we obtain the equation which describes the propagation of a second sound in medium. The obtained equation incorporates in natural way the finite velocity of heat propagation. The proposed equation is obtained as a consequence of non-linear heat transfer and is not the result of adding the second derivative term to the Fourier equation In our case that new term describes memory of the medium

The solution of the new equation describes, in addition to the well-known heat diffusion, also the ballistic propagation of the temperature pulse.

References

[1] J. Marciak-Kozlowska, M.Kozlowski, *From quarks to bulk matter*, Hadronic Press, USA, 2001

[2] Ch Kittel, H Kroemer, *Thermal physics*, Freeman, USA,1980

[3] D. Pines,P. Nozieres,*The theory of quantum liquids*,Benjamin,USA,1966